\documentclass[iicol, sn-basic]{sn-jnl}%

\usepackage{graphicx}%
\usepackage{multirow}%
\usepackage{amsmath,amssymb,amsfonts}%
\usepackage{amsthm}%
\usepackage{mathrsfs}%
\usepackage[title]{appendix}%
\usepackage{xcolor}%
\usepackage{textcomp}%
\usepackage{manyfoot}%
\usepackage{booktabs}%
\usepackage{algorithm}%
\usepackage{algorithmicx}%
\usepackage{algpseudocode}%
\usepackage{listings}%
\usepackage{url}
\usepackage{hyperref}
\usepackage{makecell}

\theoremstyle{thmstyleone}%

\theoremstyle{thmstyletwo}%

\theoremstyle{thmstylethree}%

\begin{document}

\title[Article Title]{Fast and Accurate Bayesian Optimization with Pre-trained Transformers for Constrained Engineering Problems}

\author*[1]{\fnm{Rosen (Ting-Ying)} \sur{Yu}}\email{rosenyu@mit.edu}

\author[1]{\fnm{Cyril} \sur{Picard}}\email{cyrilp@mit.edu}

\author[1]{\fnm{Faez} \sur{Ahmed}}\email{faez@mit.edu}

\affil*[1]{\orgdiv{Department of Mechanical Engineering}, \orgname{Massachusetts Institute of Technology}, \orgaddress{\street{77 Massachusetts Ave}, \city{Cambridge}, \postcode{02139}, \state{MA}, \country{USA}}}

\abstract{
Bayesian Optimization (BO) is a foundational strategy in the field of engineering design optimization for efficiently handling black-box functions with many constraints and expensive evaluations.
This paper introduces a fast and accurate BO framework that leverages Pre-trained Transformers for Bayesian Optimization (PFN4sBO) to address constrained optimization problems in engineering. Unlike traditional BO methods that rely heavily on Gaussian Processes (GPs), our approach utilizes Prior-data Fitted Networks (PFNs), a type of pre-trained transformer, to infer constraints and optimal solutions without requiring any iterative retraining. 
We demonstrate the effectiveness of PFN-based BO through a comprehensive benchmark consisting of fifteen test problems, encompassing synthetic, structural, and engineering design challenges. Our findings reveal that PFN-based BO significantly outperforms Constrained Expected Improvement and Penalty-based GP methods by an order of magnitude in speed while also outperforming them in accuracy in identifying feasible, optimal solutions. 
This work showcases the potential of integrating machine learning with optimization techniques in solving complex engineering challenges, heralding a significant leap forward for optimization methodologies, opening up the path to using PFN-based BO to solve other challenging problems, such as enabling user-guided interactive BO, adaptive experiment design, or multi-objective design optimization.
Additionally, we establish a benchmark for evaluating BO algorithms in engineering design, offering a robust platform for future research and development in the field. This benchmark framework for evaluating new BO algorithms in engineering design will be published at \url{https://github.com/rosenyu304/BOEngineeringBenchmark}.
}


\maketitle

\section{Introduction}\label{sec1}

Black-box optimization is a prevalent approach in engineering design optimization, particularly when dealing with problems where the objective function or constraints defining the set are unknown or ambiguous. This method is instrumental in navigating complex design spaces by optimizing solutions without a clear understanding of the underlying functions~\citep{bajaj2021black, alarie2021two, tao2021multi}. Lately, Bayesian optimization (BO) has emerged as a widely adopted black-box optimization tool for its ability to enable evaluations of functions with rapid speed. By leveraging probabilistic models and iteratively selecting points for evaluation, BO efficiently explores and exploits the design space, making it the forefront of active sampling for optimization~\citep{eriksson2021scalable, garnett2023bayesian, shahriari2015taking, du2023radial}. 

Though BO has emerged as a promising tool for accelerating the search process, the full realization of its potential in engineering design is hindered by the constraint-handling ability of optimization algorithms~\citep{greenhill2020bayesian, cardoso2024constrained}. A common challenge in design involves identifying products that respect all constraints in the feasible design space. Particularly in structural design optimization, constraints such as cost limitations, regulatory requirements, material constraints, geometric considerations, manufacturing limitations, safety criteria, and ergonomic factors complicate the exploration of design landscapes~\citep{gardner2017discovering, baptista2018bayesian, mathern2021multi}. Therefore, there is an increase in attention on constraint-handling BO (CBO) algorithms in both engineering and computer science communities~\citep{eriksson2021scalable, biswas2021approach, kamrah2023diverse, gardner2014bayesian, ragueneau2024constrained, tran2022aphbo, ghoreishi2019multi, tran2019constrained}.

BO's algorithm limitations also come from its commonly used surrogate: the Gaussian Process (GP). By modeling a function with the mean and the kernel (covariance) function, GP suffers from cubic time complexity $\mathcal{O}(n^3)$ of $n$ training points, leading to scalability issues and runtime concerns~\citep{liu2020gaussian, gilboa2013scaling}. The need to repeatedly refit and infer in GP-based BO exacerbates the computational time demands. Moreover, the conventional approach for handling constraints in BO, the constrained expected improvement (CEI) method~\citep{gelbart2014bayesian}, requires a separate GP for each constraint, making the time scale with the number of constraints. Therefore, studies have focused on accelerating GP-based BO and improving GP's scalability GP~\citep{cunningham2008fast, foreman2017fast, klein2017fast, martinez2018funneled, pleiss2020fast}.

To address the runtime limitations, \citet{muller2023pfns} have proposed a novel zero-training transformer framework known as Prior-data Fitted Networks for BO (PFNs4BO), which bypasses the fitting phase by leveraging a pre-trained model.  The center of this framework is the Prior-data Fitted Network (PFN), a transformer architecture meta-trained on a vast dataset of synthetically generated priors. This pre-trained nature enables the PFN to produce posterior predictive distributions without additional training, offering a promising solution for BO by speeding up the optimization process by an order of magnitude. Although PFNs4BO has demonstrated a substantial speed improvement, its applicability has been confined to single-objective optimization problems without constraints.

Despite the computer science and engineering community's numerous novel BO algorithms, there is a scarcity of open-source benchmark test suites, particularly for constrained engineering optimization problems. Various studies assess BO algorithms on distinct problem scenarios, such as material discovery~\citep{liang2021benchmarking} or chemical engineering experiments~\citep{shields2021bayesian}, making it challenging to compare methods across a broad range of engineering applications. Common optimization benchmark sets include COCO~\citep{hansen2021coco}, special sessions of competition in optimization at the Congress on Evolutionary Computation (IEEE CEC) and the Genetic and Evolutionary Computation Conference (GECCO), and Pymoo~\citep{blank2020pymoo}. However, most of these benchmarks consist of synthetic numerical problems that may not accurately represent the challenges of engineering design problems~\citep{picard2021realistic}. Of the published engineering design problems, few have their code publicly available while others are not easily interfaced with state-of-the-art BO libraries, limiting their accessibility and utility~\citep{gandomi2011mixed, yang2012bat, eriksson2021scalable, campbell2024constraining}. Additionally, most constrained engineering optimization problems in the literature are tested with Genetic Algorithms (GA)~\citep{gandomi2011mixed, yang2012bat}, making it difficult for researchers interested in testing with BO to find relevant studies. Overall, this results in a gap in the availability of a comprehensive engineering benchmark.

The focus of our study is to evaluate the performance of the recently published Prior-data Fitted Network (PFN)-based Bayesian Optimization (BO) in solving constrained engineering design optimization problems. We aim to demonstrate the effectiveness and potential superiority of general-purpose models that eliminate the need for fitting at each BO iteration, compared to traditional GP-based BO using common constraint-handling approaches. This could expand BO's applicability to time-sensitive engineering tasks such as interactive experiment design assistance or robotics control optimization. Our contributions include:

\begin{enumerate}
    \item A pre-trained transformer-based CBO algorithm: We developed a constrained-handling PFN-based algorithm utilizing the Constrained Expected Improvement (CEI) acquisition function, and which requires only one surrogate model, solving the objective and constraints in a single forward pass. PFN-CEI exhibits superior optimization performance compared to all other tested methods and is faster than GP-based CEI.
    \item Speed and performance comparison of CBO methods: We present three constraint-handling methods and two surrogate modeling approaches, evaluating their speed and optimization performance. We highlight that using PFN as the surrogate for CBO can achieve a \textbf{ten-fold increase in speed compared to GP} and outperform GP with a superior anytime performance and feasibility rate. Additionally, we provide a reflective discussion on the potential of PFN-based BO as a fast optimization algorithm. 
    \item Open-source code and enhanced benchmark tools: We present a set of fifteen test problems for benchmarking BO algorithms, featuring high-dimensional constrained engineering problems from the literature. To foster collaborative progress, we make our constrained test problem set and corresponding Python codebase available at \url{https://github.com/rosenyu304/BOEngineeringBenchmark}, encouraging other researchers to build upon and advance engineering Bayesian optimization. %

\end{enumerate}

In this work, the background of engineering design optimization problems, Bayesian optimization with constraint-handling methods, and PFN's application on BO are described in Section 2. Section 3 defines the constraint-handling BO algorithms of interest, the test problem set for benchmark, and the evaluation methods for CBO algorithms. The results of algorithm runtime and optimization performance are presented in Section 4. Finally, Section 5 discusses the overall performance of the tested CBO algorithms.

\section{Background}\label{sec2}
In this section, we introduce the common design optimization problems and methods, detail the Bayesian optimization algorithm, and highlight how the PFN-based BO method differs from traditional BO methods.

\subsection{Bayesian Optimization for Design Optimization Problems}\label{subsec2.1}
Engineering design optimization problems are often formulated as one optimization objective subjected to many inequality constraints. The mathematical representation of such problems can be written in this form:
\begin{equation}\label{eq:opt_prob}
\begin{gathered}
    \min_{x \in \mathbb{R}^d}  \quad f(x) \\
    \text{s.t} \quad g_i(x) \leq 0 \quad, i\in[1,G]
\end{gathered}
\end{equation}
where $x$ is the design variable with dimension $d$, $f(x)$ is the objective function, and $g_i(x)$ is the constraint with $G$ as the numbers of constraints.

In general, finding the optimum of an optimization problem is non-trivial.
The difficulty primarily stems from the ambiguity of objectives and constraints, coupled with the complexity of their evaluation. In engineering design, evaluating objective functions and constraints often involves physical experiments or complex simulations that are time-consuming and expensive. For instance, based on Ford Motor Company, conducting a car crash simulation may require 36 to 160 hours per experiment~\citep{wang2006review}. Researchers must then update their dataset, run the optimization algorithm—a process that itself takes time—and repeat this cycle numerous times. This leads to slow data collection and a dataset too small for accurate predictions using machine learning-based surrogate models. Thus, BO, with its efficiency in data usage and ability to incorporate prior knowledge for surrogate-based global optimization, emerges as an ideal solution for engineering design optimization tasks.

BO is an active-learning algorithm designed for black-box optimization that iteratively improves performance through exploitation and exploration ~\citep{garnett2023bayesian}. 
This process begins with a relatively small set of initial samples, typically ranging from 20 to 50 samples, depending on the complexity and dimensions of the problem at hand. BO employs a probabilistic surrogate model, commonly a Gaussian Process, to form a posterior belief about the design space. Utilizing this posterior, an acquisition function is then applied to determine the most promising next candidate, the one that is likely to be the optimum within the given space. Several acquisition functions commonly used in the literature are the probability of improvement (PI), expected improvement (EI), entropy search (ES), and upper confidence bound (UCB) ~\citep{garnett2015lecture12}. Algorithm~\ref{algoBayesianOptimization} demonstrates a general framework of the BO algorithm. 

\begin{algorithm}
\caption{Bayesian optimization (BO)}\label{algoBayesianOptimization}
\begin{algorithmic}[1]
\Require { $\mathrm{x_0}$ initial samples, $\mathrm{f(\cdot)}$ the objective function, $\mathrm{N_{iter}}$ the iterations set for the algorithm to run, $\Call{NextEval}{}$ an algorithm using: $\mathrm{Model}(\mathrm{\cdot})$ a surrogate and $\alpha$ an acquisition function for determining the next candidate for searching} 
\Ensure {The optimal solution $\mathrm{x_{opt}} $ }
\Function{BO}{$\mathrm{x_0}, \mathrm{N_{iter}}, \mathrm{f(x)}, \alpha$ }
\State Provide or perform initial sampling of $\mathrm{x_0}$
\State $\mathrm{X} \gets \mathrm{x_0}$

\For {$\mathrm{N_{iter}}$ iterations}
    \State $D \gets \mathrm{ \{X,\mathrm{f(\mathrm{X})}\} }$  %
    \State $\mathrm{x_{next}} \gets  \Call{NextEval}{\mathrm{Model}(\cdot), D, \alpha}$
    \State $\mathrm{X} \gets \mathrm{X}\cup \mathrm{x_{next}}$
    \Comment{Append $\mathrm{x_{next}}$ to $\mathrm{X}$}

\EndFor
\State \Return $ \mathrm{x_{opt}} \gets  \arg\max(\mathrm{f(X)})$
\EndFunction

\end{algorithmic}
\end{algorithm}

\subsection{Constraint-Handling Bayesian Optimization (CBO)}\label{subsec2.2} Constraint-handling Bayesian optimization (CBO) has emerged as a key area in design optimization, addressing engineering limitations such as cost, ergonomics, safety, and regulatory standards. This study focuses on solving single-objective optimization problems with $G$ constraints. Two main categories of constraint-handling approaches are typically employed for this type of problem: objective transformation and acquisition function modification.

\subsubsection{Objective transformation} \label{subsec.objectiveTransform}
One common approach for constraint handling involves penalizing the objective value of infeasible data or increasing the objective value of feasible data through an objective function transformation. The penalty function (PF) is a widely-used method that alters the objective values of infeasible data~\citep{fletcher1975ideal} by introducing a penalty term. Following this transformation, the BO algorithm is applied to the modified unconstrained optimization problem, aiming to minimize $f_{PF}$. In constrained optimization problems, the quadratic form of the penalty function is often employed, as illustrated in the following equation:
\begin{equation}\label{eq:pf}
        f_{PF}(x) = f(x) + \rho \sum^G_{i=1} max(0, g_i(x))^2\\
\end{equation}
Equation~\eqref{eq:pf} shows that, given one objective and $G$ constraints to be optimized, $f_{PF}(x)$ is the penalty transformed objective that is calculated using the objective function $f(x)$, the constraint functions $g_i(x), i \in [1,G]$, and $\rho$ is the penalty factor. The selection of the penalty factor value varies across different studies. In this paper, we initialize $\rho=1$ and multiply $\rho$ by 1.5 if the algorithm fails to identify an improved optimal value after five iterations~\citep{campbell2024constraining}. However, a limitation of this method is its difficulty in implementation when the constraints cannot be represented analytically by numerical equations.

\subsubsection{Acquisition Function Modification} \label{subsec.AcqCEI} As evaluating black-box constraints alongside the objective function through analytical transformation can be challenging, one strategy is to treat constraint functions as feasibility objectives. This requires surrogate modeling and their inclusion in the acquisition function calculation. Consequently, the constraint-handling Bayesian optimization (CBO) process employs $1+G$ surrogate models for modeling the objective function with $G$ constraints. These surrogate models for constraints, known as feasibility models, determine the probability of feasibility $P_{feas}$ as shown in Equation~\eqref{eq_pfeas}.

One of the most popular objective acquisition functions, expected improvement (EI, see Equation~\eqref{eq_EI}), has been adapted for constrained optimization. ~\citet{gelbart2014bayesian} proposed the constrained EI (CEI) acquisition function as the sum of EI and $P_{feas}$ for each constraint shown in Equation~\eqref{eq_cei}.
\begin{equation} \label{eq_pfeas}
    P_{feas} = \Phi \left( \frac{- \hat{g}(x) }{\sigma_{g(x)}} \right) 
\end{equation} 
\begin{align} \label{eq_EI}
\begin{split} 
        \alpha_{EI} =& \left(f_{*} - \hat{f}(x)\right)  \Phi \left( \frac{f_{*} - \hat{f}(x) }{\sigma_{f(x)}} \right) \\
        &+ \sigma_{f(x)} \mathcal{N} \left( \frac{f_{*} - \hat{f}(x) }{\sigma_{f(x)}} \right) 
\end{split} 
\end{align}
\begin{gather} \label{eq_cei}
        \alpha_{CEI} =  \alpha_{EI}  \prod_{i=1}^G  P_{feas, i} 
\end{gather} 
where $\Phi$ is the Gaussian cumulative distribution function (CDF), $\hat{g}(x)$ is the mean value of $g$ at point $x$, $\sigma_{g(x)}$ is the standard deviation of $g(x)$, $f_{*}$ is the minimum (optimum) observed value up until the current iteration, $\hat{f}(x)$ is the mean value of $f$ at point $x$, $\sigma_{f(x)}$ is the standard deviation of $f(x)$, and $\mathcal{N}$ is the Gaussian (Normal) distribution.

One limitation of CEI is that the combined probability of feasibility will be close to zero at the edge of the constraint regions. Therefore, a modified version of the CEI algorithm has been proposed to increase the chance of selecting solutions near the constraint boundaries~\citep{bagheri2017constraint} :

\begin{gather} \label{eq_cei+}
        \alpha_{CEI+} = \alpha_{EI} \prod_{i=1}^G min(1,2P_{feas, i} ) 
\end{gather}

\subsection{Prior-Data Fitted Network and its application on Bayesian Optimization}\label{subsec2.3}
A Prior-Data Fitted Network (PFN) is a transformer framework trained to perform Bayesian inference~\citep{muller2021transformers}. Unlike conventional surrogate models trained on a single dataset and must be retrained when new data is observed, PFN is designed to be trained only \textbf{once}. After this one-time meta-training, PFN uses its encoder-only transformer structure during inference to compute the posterior predictive distribution (PPD) $p(y|x, D)$, where $x$ denotes the input samples, $y=f(x)$ the response, and $D = \{(x_1, y_1), ..., (x_k, y_k)\}$ the observed dataset. The samples are passed through a ``frozen'' model as it is used at the inference time, meaning its parameters, weights, and biases are kept fixed. PFN employs the attention mechanism that conditions on known sample D, similar to how Large Language Models (LLMs) like ChatGPT are conditioned by text prompts but without position encoding. This mechanism enables PFN to utilize the most relevant prior information for new problems and make predictions for unlabeled samples.

PFNs are trained on a vast and varied dataset with millions of prior data points during meta-learning to `learn' the execution of general tasks. In the context of BO, PFN is trained to mimic GP and is trained on well-designed prior data inspired by HEBO Bayesian Optimization solver~\citep{cowen2022hebo}. This prior data incorporates non-linear input and output warping, enhancing the robustness of surrogate modeling. Additionally, PFNs extend HEBO by incorporating a well-engineered GP prior, making them highly effective at capturing complex data dependencies.

Specifically, the PFNs4BO framework involves using PFN as a surrogate for approximating PPD and an acquisition function for getting the next search point in optimization. At every iteration, the transformer model simultaneously processes the known data $D=\mathrm{ \{X_{n\times d},\mathrm{f(\mathrm{X})_{n\times 1}}\} }$ and the pending search points  $\mathrm{X_{pending_{m\times d}}}$, where $m \gg n$, simultaneously to calculate the acquisition value for $\mathrm{X_{pending}}$. 
This single-pass process is similar to the fitting phase of GPs and also the ``training'' and ``testing'' of neural networks, yet it involves no real training or fitting as it is used at inference time. Furthermore, PFN's capability to perform posterior prediction on large size of $\mathrm{X_{pen}}$ makes it highly effective for exploring the search space even without an acquisition function optimizer. The output posterior predictive distribution is then fed into an acquisition function to determine the next search point with the maximum acquisition value. Algorithm \ref{alg:NextEval} emphasizes the differences between GP-based BO and PFN-based BO. GP-based BO requires refitting at every iteration and the use of an acquisition function optimizer, while PFN-based BO does not require either of these.

There are also limitations to PFNs. The current released PFN model on the PFNs4BO GitHub repository\footnote{\url{https://github.com/automl/PFNs4BO/}} can only take data up to 18 design variables. To accommodate design problems with higher dimensions, retraining is needed to generate a larger PFN, and the training time is less than 24 hours on a cluster node with eight RTX 2080 Ti GPUs~\citep{muller2023pfns}. Additionally, the PFNs4BO framework can only do a single objective optimization problem with three acquisition functions: EI, PF, and UCB. The current framework does not have any capability to handle constraints.
In this study, we address this gap and add a constraint-handling acquisition function by exploiting PFN's transformer nature to pass and solve data in parallel.

\begin{algorithm}
\caption{Surrogate modeling and acquisition function}\label{alg:NextEval}
\begin{algorithmic} [1]
\Require{$\mathrm{Model}(\cdot)$ that returns a posterior distribution of the input data 
$D$, $\alpha$ an acquisition function, $\mathcal{X}$ search space}
\Ensure{$\mathrm{x_{next}}$ the next search point} 
\Function{NextEval}{$\mathrm{Model}(\cdot)$, D, $\alpha$ }
\If {$\mathrm{Model}$ is $\mathbf{GP}$}
    \State $\mathbf{GP}(D)$ $\gets$ Fitting $\mathbf{GP}$ with D 
    \State $\mathrm{x_{next}} \gets \underset{\hat{x}\in\mathcal{X}}{\arg\max}  \  \alpha ( \hat{x}, \mathbf{GP}(D) )$ \\
    \Comment{optimizing $\alpha$}
\EndIf
\\
\If {$\mathrm{Model}$ is $\mathbf{PFN}$}
    \State $\mathrm{x_{next}} \gets \underset{\hat{x}\in\mathcal{X}}{\arg\max}  \ \alpha ( \hat{x}, \mathbf{PFN_{\theta}} (\cdot | D ) )$ 
\EndIf

\State \Return $\mathrm{x_{next}}$
\EndFunction
\end{algorithmic}
\end{algorithm}

\section{PFN-based CBO Frameworks}
\label{section:PFNCBO}
Here we propose three PFN-based CBO frameworks with three different constrain-handling approaches: PFN-Pen (PFN with penalty function), PFN-CEI (PFN with constrained EI), and PFN-CEI+ (PFN with modified constrained EI). Figure~\ref{fig:Algorithm} visualizes the difference between GP-based and PFN-based BO.
\subsection{PFN-Pen} \label{sec:PFNPen} Using the penalty transform method discussed in Section \ref{subsec.objectiveTransform}, PFN-Pen performs Bayesian optimization on the transformed objective $f_{PF}(X)$ and outputs the posterior for acquisition function $\alpha_{EI}$. For calculating $f_{PF}(X)$, we initialize the penalty factor $\rho=1$ and multiply $\rho$ by 1.5 when the algorithm fails to identify an improved optimal value after five iterations \citep{jetton2023constrained}.

    \begin{gather}
        \alpha_{EI} ( \mathbf {PFN_{\theta}} (\cdot | \{X,f_{PF}(X)\}  )
    \end{gather}

\subsection{PFN-CEI} \label{sec:PFNCEI} 
\begin{figure*}[htbp]
	\centering
 \includegraphics[width=1\linewidth]{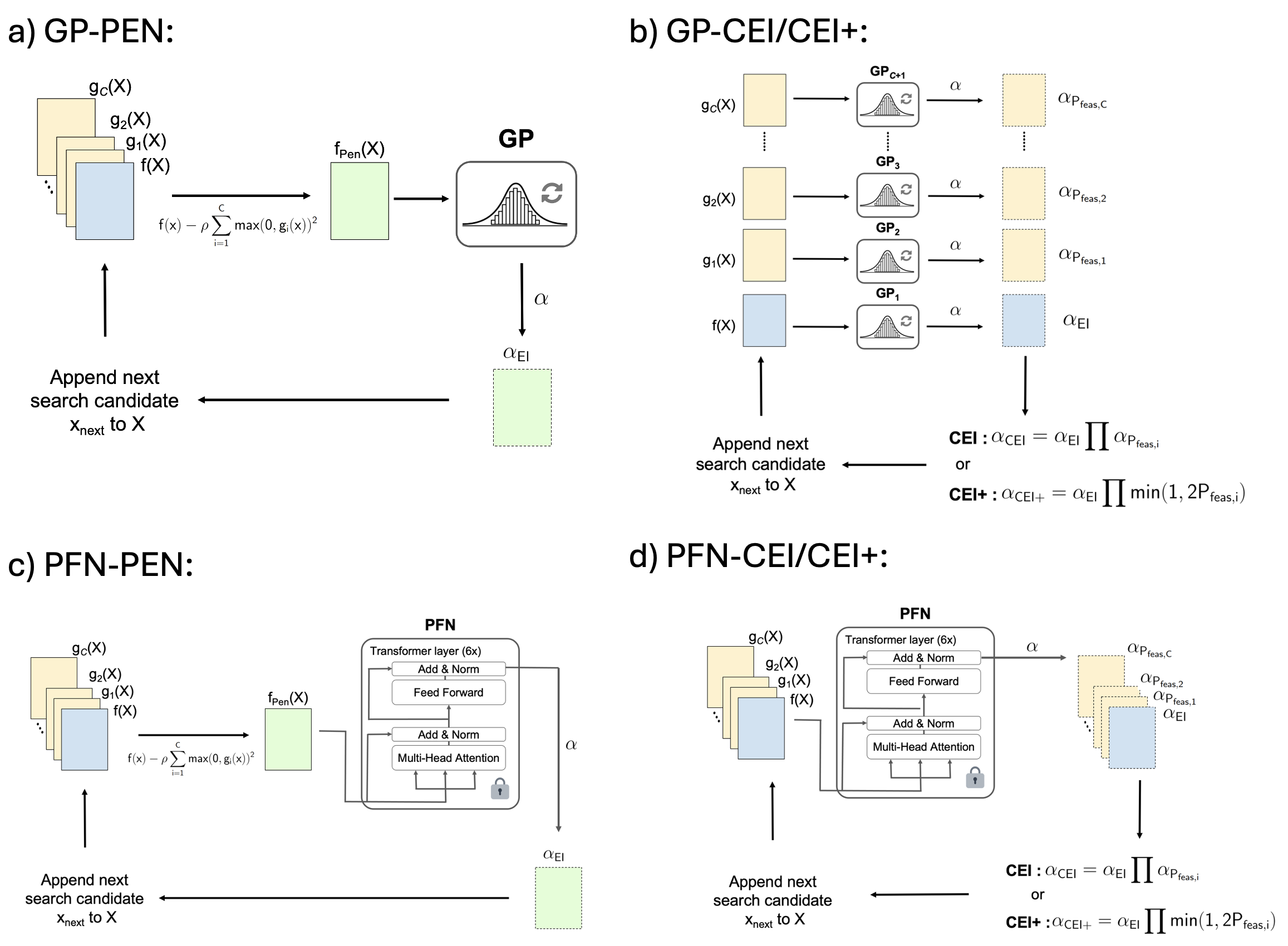}
 \caption{(a) GP-Pen; (b)GP-CEI/CEI+: Given an objective and $G$ constraints, GP-CEI will need $G+1$ GPs to perform one search iteration for BO. Each GP will be fit and updated in every iteration; (c) PFN-Pen; (d) PFN-CEI/CEI+: Only one PFN is needed for optimizing an objective and $G$ constraints, and no fitting of PFN will happen during BO since it is a pre-trained model. PFN's transformer nature allows the EI of the objective and $P_{feas}$ of the constraints to be solved in parallel in one pass. }

    \label{fig:Algorithm}
\end{figure*}

To implement CEI constraint-handling method stated in Section~\ref{subsec.AcqCEI}, the calculation of $\alpha_{EI}$ and $P_{feas}$ of each constraint function are required. In contrast to the GP-based approach, which requires a separate GP for each objective and constraint, a PFN can solve for the acquisition values for $\alpha_{EI}$ and $P_{feas}$ in one forward pass using a single surrogate. Leveraging the transformer architecture of PFN, which supports batch processing, we develop a method to simultaneously solve objectives and constraints with a single model. Figure~\ref{fig:Algorithm}'s (b) and (d) highlight the differences between GP-CEI and PFN-CEI.

    \begin{gather}
        \alpha_{CEI} ( \mathbf {PFN_{\theta}} (\cdot | \{X,f(X)\} ) )
    \end{gather}

\subsection{PFN-CEI+} \label{sec:PFNCEI+} This method adds the modified CEI+ threshold mentioned in Section~\ref{subsec.AcqCEI} to the PFN-CEI algorithm to handle constraint boundaries.

    \begin{gather}
        \alpha_{CEI+} ( \mathbf {PFN_{\theta}} (\cdot | \{X,f(X)\} ) )
    \end{gather}

\section{Experiments} \label{section.Method}

This section describes our constrained optimization problems, on which constrain-handling Bayesian Optimization algorithms are tested, and the evaluation metrics.

\subsection{CBO Algorithms}\label{subsec3.2} In this study, we focus on benchmarking the performance of three proposed PFN-based CBO algorithms as highlighted in Section~\ref{section:PFNCBO} with the current state-of-the-art GP-based BO using BoTorch library~\citep{balandat2020botorch}, an open-source Bayesian optimization tool based on PyTorch. To make a fair comparison between algorithms using two different surrogates, we implement the same constraint-handling methods on GP and formulate three GP-based CBO algorithms: GP-Pen, GP-CEI, and GP-CEI+. A detailed visualization of all six CBO algorithms tested in this study is detailed in Figure~\ref{fig:Algorithm}.

\subsection{Test Problems}\label{subsec3.1} 
\begin{figure*}[htbp]
	\centering
 \includegraphics[width=1\linewidth]{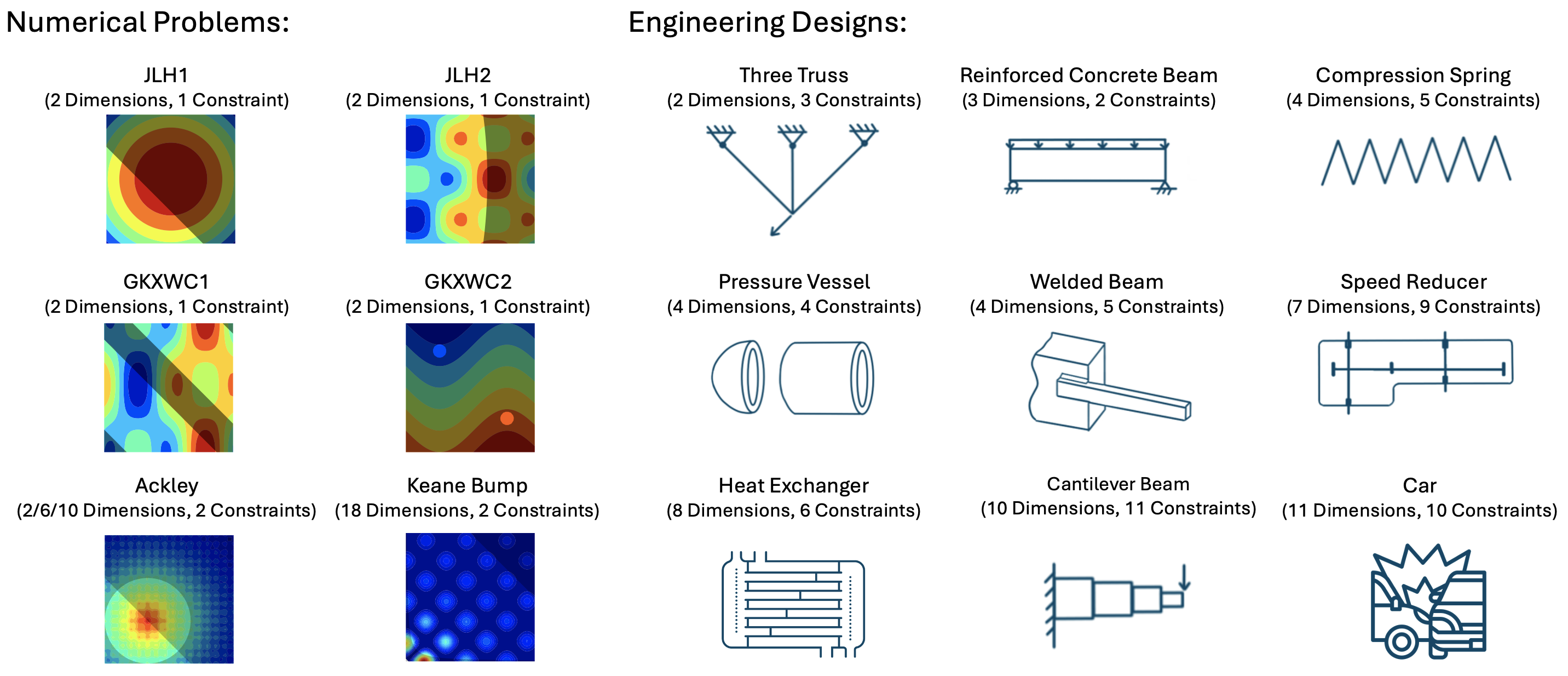}
    \caption{ Overview of the details of the 15 benchmark problems. The non-feasible regions are shaded in the numerical problems. The Ackley problem is experimented with optimization in 2D, 6D, and 10D, resulting in 17 experimental trials. }
    \label{fig:icon}
\end{figure*}

This study incorporates a diverse set of constrained test problems gathered from the literature of structural optimization algorithms ~\citep{gandomi2011mixed, koziel2011computational, yang2012bat, jetton2023constrained}. With a focus on benchmarking the algorithm's ability to solve engineering problems, we gather six numerical test problems and nine engineering design optimization problems with both continuous and discrete value optimization. These fifteen problems are detailed in Figure~\ref{fig:icon} and Appendix A.   

\subsection{Algorithm Tests} \label{subsec:AlgorithmTests}

The goal of the algorithm test is to provide a fair and comprehensive comparison of the state-of-the-art optimization methods. In this research, the optimization goal is to minimize the objective function for the test problems. The initial samplings for all test problems are performed with Latin Hypercube Sampling. Each test problem includes fifty sets of initial samples that are randomly selected, with each initial set representing a separate experimental trial. In each individual experimental trial, all six algorithms begin the optimization process with the same set of initial samples. Furthermore, each algorithm is run for 200 iterations of optimization and timed for the run time for each experimental trial. The optimal searched value and the total run time of the fifty trials for each test problem are then evaluated with our ranking procedure.

All algorithms are run on the same computer to ensure the speed comparison is fair. The CPU is Intel® Core™ i9-13900K Processor with 24 cores, and the RAM has 128GB. The system is GNU/Linux 6.5.0-15-generic x86\_64 with Ubuntu 22.04.3 LTS as the operating system. While transformer models gain significantly from GPU-acceleration and parallelization, we use only CPUs for a fair comparison with other methods.

\subsection{Evaluation metrics} \label{subsec:AlgorithmEvaluation}
\subsubsection{Feasibility Ratio} \label{subsec:FeasibilityRatio} 
We define the optimization solution as the minimal value found by each method during optimization. However, for the BO algorithms utilized in this study, there is no guarantee of convergence to a feasible solution that respects all constraints. Therefore, the feasibility of the solution generated by each method is utilized as a metric for method evaluations. We define our constraint-handling performance evaluation methods as:
\begin{equation}
        \text{Feasibility ratio} = 
        \frac{\text{\# trials with feasible solution }}{\text{Total \# trials = }50}
\end{equation}

\subsubsection{Statistical Ranking } \label{subsec:StatisticalRanking} A statistical ranking evaluation is used to evaluate the performance of our BO algorithms. We analyze two features of each BO method: the minimization result, and the algorithm run time for performing 200 iterations of the BO search. The ranking approach we used is widely used for ranking machine learning algorithms~\citep{IsmailFawaz2018deep}.

The ranking process first has the BO methods ranked based on their performance for each problem, with the best-performing method receiving the lowest rank. The Friedman ranking test is then applied to the ranks to reject identical methods with a threshold p-value set at 0.05. Next, for the hypothesis testing, Wilcoxon significance analysis and Holm’s adjustment are performed to compare the algorithms pairwise. Based on the adjusted p-values, this statistical approach distinguishes the methods that are significantly different from each other. For details about the ranking process, see the original paper that uses this ranking method for evaluating PFN's classification performance~\citep{picard2024fast}. 

\subsubsection{Fixed-budget Analysis} \label{subsec:FixedbudgetAnalysis} \citet{hansen2022anytime} presented the concept of fixed-budget evaluations, a technique for comparing the efficiency of optimization algorithms by allotting specific computational resources for their execution. Our investigation employs two distinct fixed-budget analysis methodologies:
\begin{enumerate}
    \item Fixed-iteration approach: The performance of each optimization algorithm is evaluated after a pre-determined number of 200 iterations.
    \item Fixed-runtime approach: The performance outcomes of the algorithms are compared within an identical CPU time frame. In this study, the runtime budget is set to be the time required for the fastest method to execute 200 iterations.
\end{enumerate}

\section{Results}\label{subsec4} 
\subsection{Feasibility Ratio Performance}\label{subsec4.1} The feasibility ratio quantifies the capability of identifying a useful solution within the constrained space after a fixed number of iterations. Table~\ref{tabFeas} presents the feasibility scores for each method across the test problems. For most test problems, all methods successfully find a feasible solution. Yet, not all algorithms can find a feasible solution for the Ackley function (2D, 6D, 10D), GKXWC2, and the Heat Exchanger problem every time. In these more challenging problems, we note that algorithms employing CEI for constraint handling exhibit a higher feasibility ratio than those utilizing a penalty function. For instance, in the Ackley 10D problem, GP-Pen achieves 32\% feasible results, while GP-CEI and GP-CEI+ reach 86\% and 78\% feasibility rates, respectively. This trend is even more evident in the Heat Exchanger example. The feasibility ratio for GP-based methods increases from 2\% to approximately 80\% with the implementation of CEI, and for PFN-based methods, it rises from 40\% to 100\% when switching from the penalty transform to CEI.

The feasibility ratio analysis also reveals that the simplest method, GP-Pen exhibits the lowest feasibility rate, as expected. For high-dimensional problems, such as Ackley10D, PFN-based constrained BO methods demonstrate a higher feasibility rate than GP-based methods overall. 

\begin{center}
\begin{table*}[htbp]
\caption{Feasible Rate of different CBO methods.}\label{tabFeas}%
\begin{tabular}{@{}lllllll@{}}
\toprule
Test case & GP-Pen  & GP-CEI  & GP-CEI+ & PFN-Pen & PFN-CEI & PFN-CEI+\\
\midrule
JLH1 & 100\%   & 100\% & 100\% & 100\%& 100\% & 100\%\\
JLH2 & 100\%   & 100\% & 100\% & 100\%& 100\% & 100\%\\
GKXWC1 & 100\%   & 100\% & 100\% & 100\%& 100\% & 100\%\\
GKXWC2 & 92\%   & 100\% & 100\% & 100\% & 100\% & 100\%\\
Ackley 2D & 98\% & 100\% & 100\% & 100\% & 100\% & 100\%\\
Ackley 6D & 100\%   & 98\% & 98\% & 100\% & 100\% & 100\% \\
Ackley 10D & 32\%   & 86\% & 78\% & 92\% & 94\% & 92\%\\
Three Truss & 100\%   & 100\% & 100\% & 100\%& 100\% & 100\%\\
Reinforced Concrete Beam & 94\%    & 100\% & 100\% & 100\%& 100\% & 100\%\\
Compression Spring & 100\%   & 100\% & 100\% & 100\%& 100\% & 100\%\\
Pressure Vessel & 100\%   & 100\% & 100\% & 100\%& 100\% & 100\%\\
Welded Beam & 100\%   & 100\% & 100\% & 100\%& 100\% & 100\%\\
Speed Reducer & 100\%   & 100\% & 100\% & 100\%& 100\% & 100\%\\
Heat Exchanger & 2\%   &  80\% & 82\% & 40\% & 100\% & 100\%\\
Cantilever Beam & 100\%   & 100\% & 100\% & 100\%& 100\% & 100\%\\
Car & 100\%   & 100\% & 100\% & 100\%& 100\% & 100\%\\
Keane Bump 18D &  100\%   & 100\% & 100\% & 100\%& 100\% & 100\%\\

\botrule
\end{tabular}

\end{table*}
\end{center}

\subsection{Optimization Performance at Fixed-iteration}\label{subsec4.2} Figure~\ref{fig:Stats} displays the distribution of optimal and feasible solutions for each method across 17 problems of 200 iterations. Our analysis begins with the optimization performance of six CBO algorithms in different categories of optimization problems. For numerical problems such as Ackley 6D and 10D, GP-CEI+ is 60\% and 68\% better than PFN-CEI in optimization performance. However, note that these represent only the feasible samples, and Table 2 shows that GP-CEI+ only generates 78\% feasible samples, while PFN-CEI generates 94\% feasible samples for Ackley 10D. The Keane Bump 18D problem is known to be challenging for GP-based methods ~\citep{eriksson2021scalable}, where the PFN-based methods surpass the GP-based methods by 10\%. 

For the nine engineering problems, PFN-based methods consistently rank highest compared to GP-based methods. The median solutions from the PFN-CEI method dominate all engineering problems, exhibiting performance two to three times better than that of GP-CEI or GP-CEI+.

\subsection{Optimization Performance at Fixed-Runtime}\label{subsec4.new} Figure~\ref{fig:Converge} illustrates the convergence plot for each problem, highlighting the optimal value at a fixed time constraint marked by the completion of 200 iterations by PFN-Pen, the fastest approach. Upon the completion of PFN-Pen, a comparative analysis of performance outcomes reveals that PFN-CEI outperforms the others in 10 of the problems, while PFN-Pen leads in 6 cases, and PFN-CEI+ prevails in 1 case. PFN-based strategies consistently exhibit superior anytime performance throughout the operational timeframe defined by the termination of PFN-Pen. 

Additionally, the convergence plot shows the advantage of PFN-based BO in limited runtime search, where GP-based BO sometimes cannot find feasible solutions. Specifically, for the Ackley 6D and 10D problems, although GP-based CEI methods outperform PFN-based methods after two hundred iterations, the GP-based method is unable to find any constrained optimal solution within the given runtime limit. Even more evident, in the Compression Spring and Heat Exchanger problems, the PFN-based method outperforms in optimization, while GP-CEIs fail to find a feasible solution in the given time budget and perform worse than PFN-CEI at the fixed iteration.

\subsection{Speed Performance}\label{subsec4.3} 
\begin{figure*}[htbp]
	\centering
 \includegraphics[width=1\linewidth]{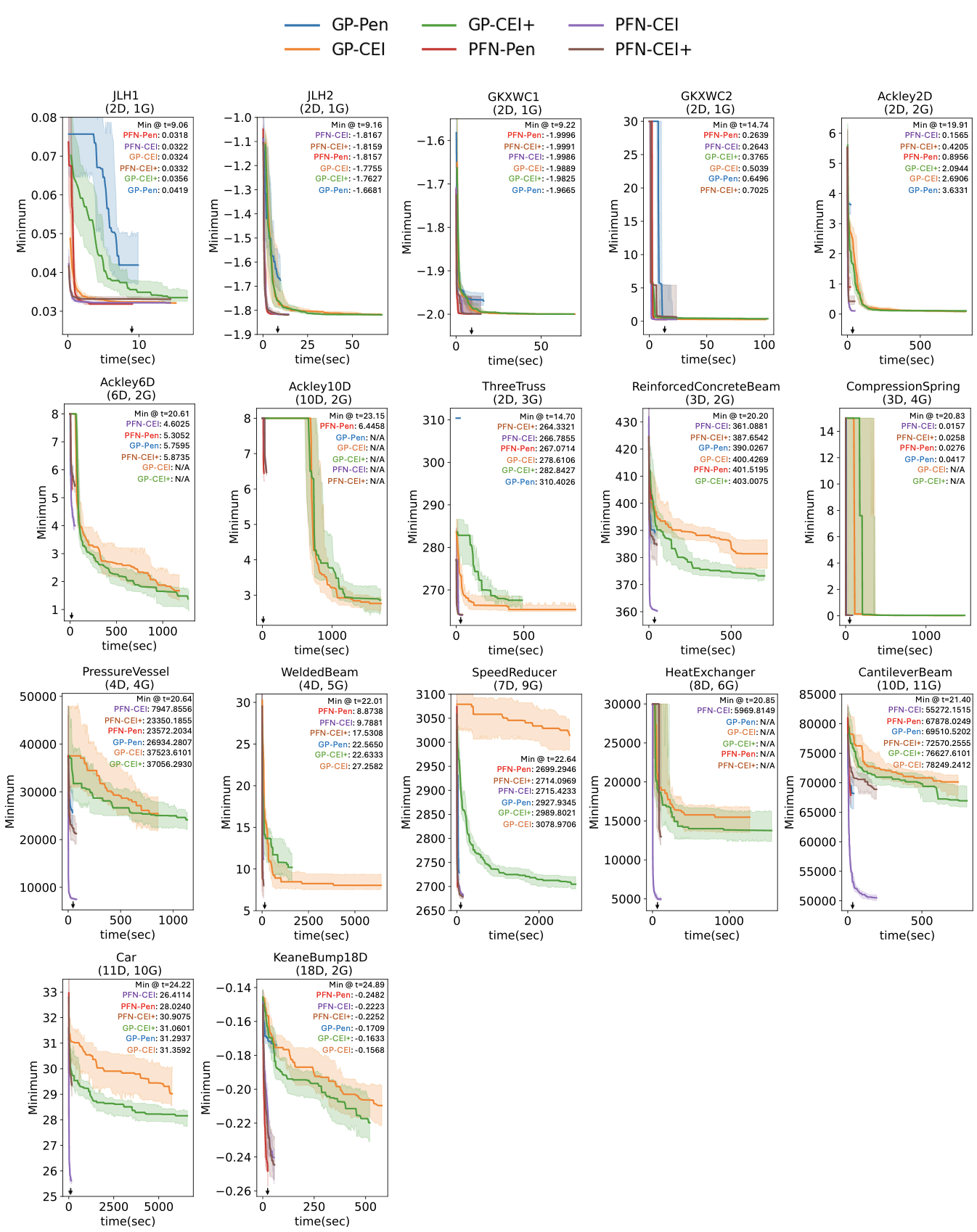}
    \caption{Convergence plots comparing the optimal value of each method for each experiment at a fixed runtime budget. The runtime budget is set to be when PFN-Pen, the fastest method, finishes running 200 iterations. The minimum value of all methods at this fixed time budget is sorted and shown in each plot, with the value at the top being the best performance.}
    \label{fig:Converge}
\end{figure*}

\begin{figure*}[htbp]
	\centering
 \includegraphics[width=1\linewidth]{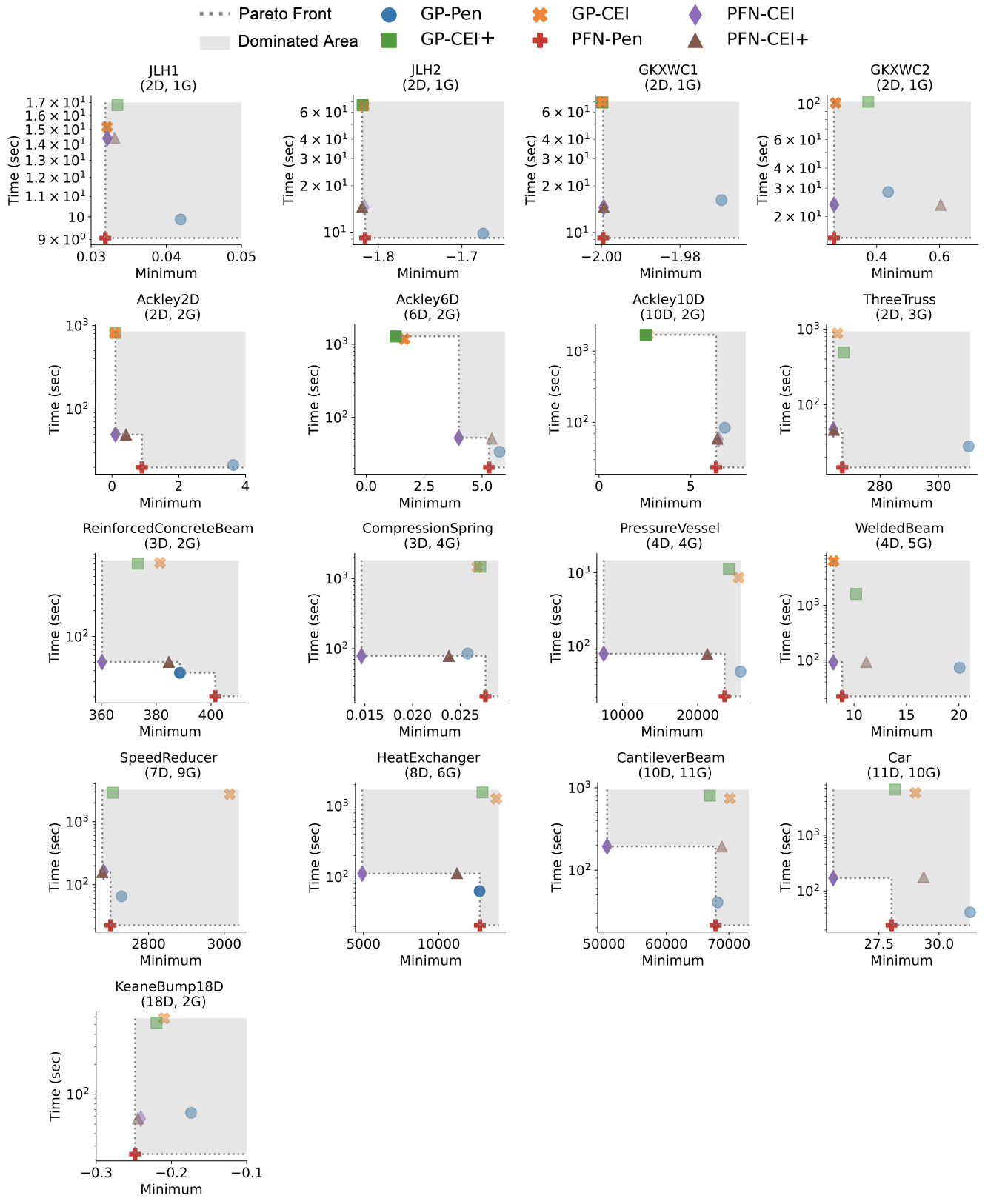}
    \caption{Pareto plots demonstrating the trade-off between performance and total execution time (log-scale) for each method and test problem. D is the objective dimension, and G is the number of constraints. The average Pareto rank of each method over seventeen experiment trials is [GP-Pen, GP-CEI, GP-CEI+, PFN-Pen, PFN-CEI, PFN-CEI+] = [2.118, 2.353, 2.353, 1, 1.353, 1.765], where the smaller rank, the better, and rank 1 is the best. In problems with more than one constraint, PFN-based methods are 10 times faster than the GP-based CEI methods. }
    \label{fig:Paerto}
\end{figure*}

In addition to the convergence plot in Figure~\ref{fig:Converge}, Figure~\ref{fig:Paerto} illustrates the Paerto trade-off between time and performance for each test problem. In both Figures, algorithms using the penalty function are always faster than those utilizing CEI. PFN-Pen leads in speed on the Pareto front in all seventeen benchmark problems, completing two hundred iterations in 17.8 seconds on average. On the other hand, GP-Pen requires 36.5 seconds since GP is affected by the problems' expanding dimensions.

The speed disadvantage of GP's approach becomes more evident when comparing CEI methods for constraint handling. GP-CEI and GP-CEI+ need 647.9 and 590.8 seconds on average to perform two hundred iterations of search, with a maximum of 6588.9 seconds (1.77 hours) for running the Heat Exchanger problem. In contrast, PFN-CEI and PFN-CEI+ only need 55.8 seconds on average to finish an experiment, showing that they are 10 times faster than the GP-based CEI methods. 

Moreover, CEI algorithm speeds drop as $G$ goes from 1 to 11, with GP-based methods taking 430 times longer because of the feasibility calculations for each constraint. Due to the PFN's capability to solve both objectives and constraints in a single forward pass, PFN-based CEI methods' speeds only drop by a factor of 13, demonstrating the dominance of PFN-based methods in speed.

\subsection{Overall Rank}\label{subsec4.4}
\begin{figure}[h!]
	\centering
 \includegraphics[width=0.9\linewidth]{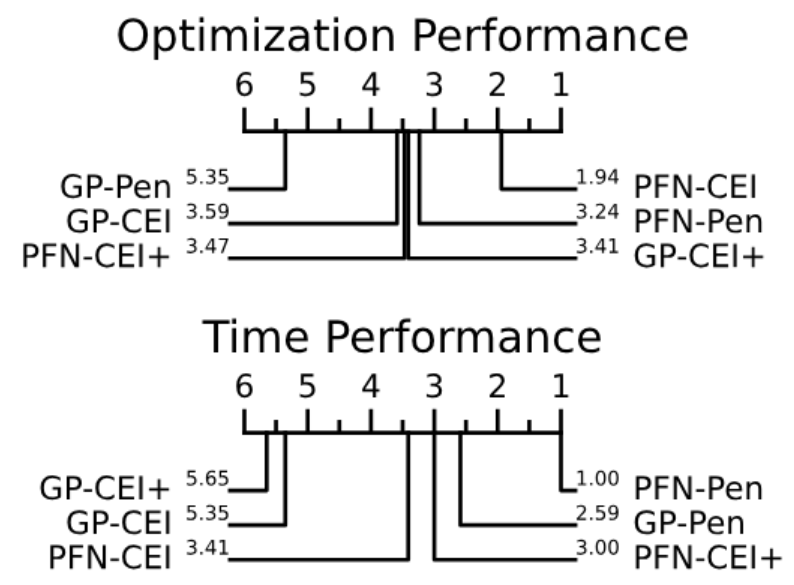}
    \caption{ Critical difference rank plot of overall results. A smaller rank indicates a better result. Regarding optimization performance, two PFN-based methods (PFN-CEI and PFN-Pen) lead. For time performance, PFN-Pen dominates as the fastest method, while the GP-based CEI ranks last. }
    \label{fig:Rank}
\end{figure}

Figure~\ref{fig:Rank} shows the critical difference plot from the statistical ranking of the six different CBO approaches. The performance result shows that two PFN-based methods, PFN-CEI and PFN-Pen, outperform the traditional GP-based methods for the optimization performance rank. Surprisingly, applying CEI+ onto PFNs does not help improve the performance as it does for GP-CEI+. Furthermore, the time results show that the two methods using the penalty transform require the least time, with the PFN-Pen being the fastest method. The GP-CEI and GP-CEI+ methods are the slowest as expected.

\begin{figure*}[htbp]
	\centering
 \includegraphics[width=1\linewidth]{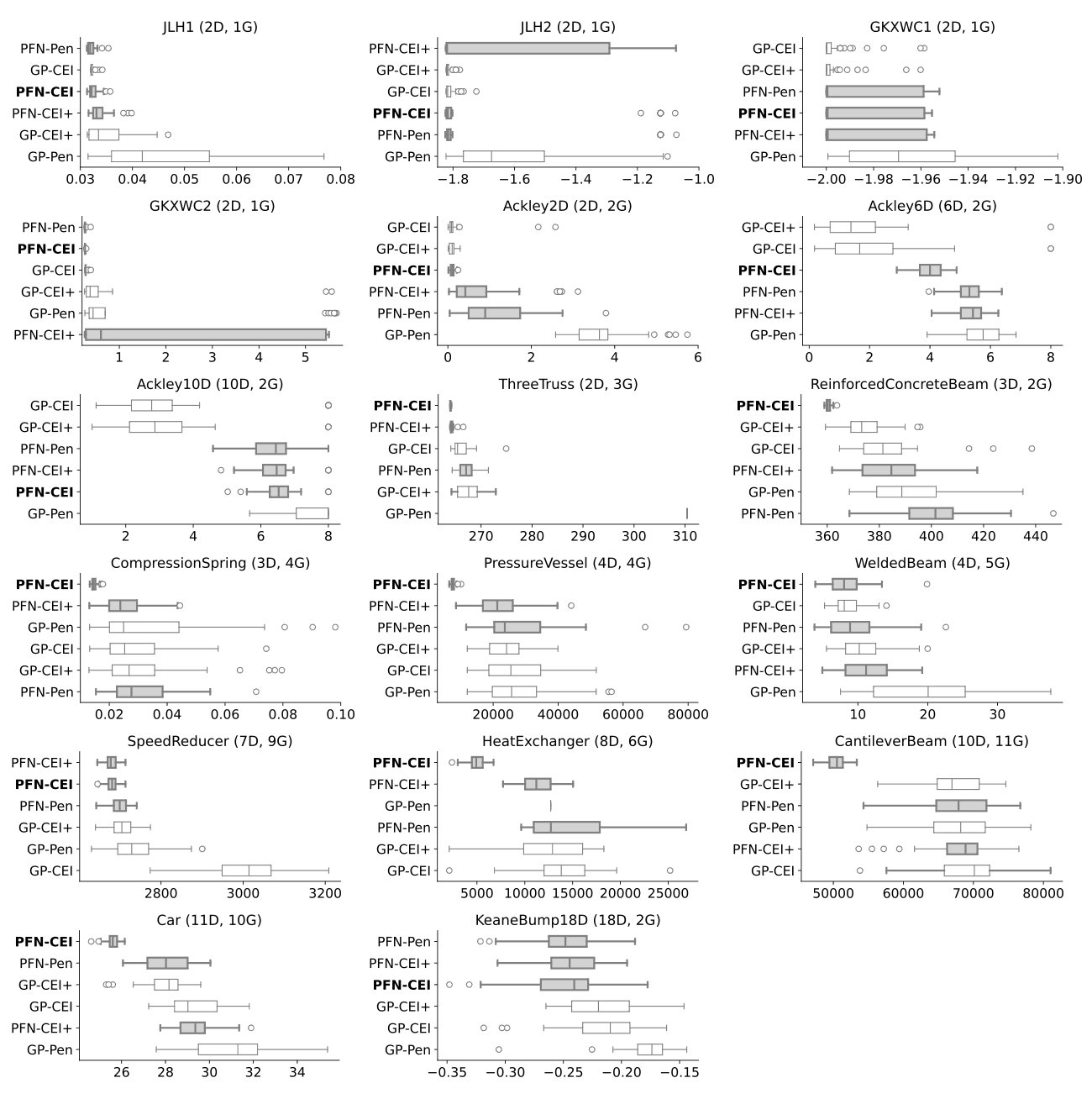}
    \caption{Box plots comparing the optimal value of each method for each experiment. }
    \label{fig:Stats}
\end{figure*}

\section{Discussion}

\subsection{Recommendations for Constrained Bayesian Optimization Methods}\label{sec5.1} Overall, the results show that using PFN as BO's surrogate outperforms GP in speed and optimization performance. Notably, PFN algorithms, already 10 times faster than GP, could further increase their speed through parallelization, as we tested all our models only on CPU. Considering the optimization performance, PFN-CEI has the top rank in optimization performance, followed by PFN-Pen. Though PFN-Pen has a faster speed for doing 200 iterations, PFN-CEI converges faster, as shown in Figure~\ref{fig:Converge}. Therefore, we recommend using PFN-CEI for overall performance and could do an early stop before 200 iterations if the user wants to perform optimization with a limited runtime budget.

\subsection{Potentials of Pre-trained Transformers-based BO}\label{sec5.PFN} From the convergence plot and fixed-runtime analysis, it is evident that pre-trained-model-based BO is effective in rapidly assessing optimization problems and providing feasible solutions when GPs are unable to do so. Observing PFN's capability for rapid optimization through the transformer's ability to solve multiple functions in parallel in a single forward pass, we want to emphasize the potential of using pre-trained models for BO and its applications. Experimental optimization or user-guided BO~\citep{campbell2024constraining}, which requires human input to the engineering optimization framework, will be time-sensitive as users must wait for BO to indicate the next potential optimum. Fast BO enables users to receive immediate feedback, enhancing the efficiency of the workflow. Conversely, BO for hyperparameter tuning has been shown to take an extended period to identify the best hyperparameters for large models in image classification or language modeling~\citep{cho2020basic}. Transformer-based BO could potentially unlock the possibilities of fast hyperparameter optimization.

\subsection{Understanding the Complexity of Test Problems}\label{sec5.2} One metric for comprehending the intricacy of the constrained test problems is to evaluate the feasible ratio of the six methods for each problem. Multi-modal numerical problems like Ackley, and problems with relatively small constrained areas, such as GKXWC2, are particularly challenging since not all methods have a 100\% feasible ratio. With higher feasibility ratios in most engineering test cases, BO proves effective for constrained engineering design problem-solving. The Heat Exchanger problem, however, demonstrates the lowest overall feasibility among all problems with the longest runtime due to the exclusive presence of independent variables in the constraints and not in the objective function, making it the most complex engineering problem. 

An alternative metric for evaluating problems is the variance in results across methods, as shown in Figure~\ref{fig:Stats}. The choice of BO method is particularly vital for numerical problems, where results vary greatly, especially in cases like JHL2, GKXWC1, GKXWC2, and Ackley, due to their complexity. Engineering problems such as Pressure Vessel and Speed Reducer also have large variances in their results, underscoring the importance of method selection.

\subsection{Limitations}\label{sec5.3} This study provides insights into the application of PFN as black-box surrogates for BO while also acknowledging several inherent limitations. Firstly, the PFN-based methods employed in this research do not utilize acquisition function optimizers that are commonly implemented in BO algorithms. Theoretically, the absence of an acquisition optimizer could potentially accelerate the algorithm but suppress the optimization performance. While PFN-based BO still had the overall best performance, adding acquisition functions to them could further enhance their performance in multi-modal problems such as Ackley. Therefore, further evaluations of PFN-based BO utilizing standard acquisition optimization approaches are required. 

Additionally, while PFNs show promise in addressing our nine engineering design problems through BO, this limited scope may not capture the full complexity and diversity encountered in practical engineering situations. We hope that our research will encourage other scholars to adopt PFN-based BO for their engineering design tasks and to evaluate novel algorithms using our benchmark problem sets. We also aim to expand the benchmark problem based on community input, providing a standardized test bed for research in constrained BO methods.

Lastly, our evaluation metrics were limited to runtime speed and optimization performance in fixed iterations. Future studies should explore additional aspects of Bayesian optimization, such as the convergence rate of iterations and scalability.

\subsection{Future Work} \label{sec5.4}
Our work in this paper lays the groundwork for several promising paths for future research. Firstly, we aim to expand our constrained PFN-based BO methods to multi-objective or active sampling PFN by introducing additional acquisition functions or modifying the PFN structure. Furthermore, we plan to benchmark our approach against more constraint BO methods such as Scalable Constrained Bayesian Optimization (SCBO) ~\citep{eriksson2021scalable}, as our current comparison is limited to CEI. This will involve a detailed comparison of PFN with other CBO methods using active sampling strategies or BO methods that employ neural networks instead of Gaussian Processes. Given the rapid computation speed of PFN, we foresee the potential for solving high-dimensionality problems with PFN models using strategies such as bootstrapping and aggregation. This expansion not only enhances the versatility of our approach but also opens up new possibilities for tackling complex optimization challenges in multidisciplinary optimization.

\section{Conclusions}\label{sec6}
This research evaluates a novel approach for constraint-handling Bayesian optimization (CBO) by utilizing prior-data fitted networks (PFN) to remove the need for re-fitting the Gaussian Process (GP) for every searching iteration. Our comprehensive analysis is supported by benchmarking the methods on the 17 constrained optimization experiments, ranging from numerical synthesized test cases to engineering design problems. By using three constraint-handling approaches, penalty function PF, constrained expected improvement CEI, and modified constrained expected improvement CEI+, and two different surrogates, we evaluate six different CBO algorithms: GP-Pen, GP-CEI, GP-CEI+, PFN-Pen, PFN-CEI, PFN-CEI. The results show that across the optimization problems, the PFN-based approach has dominated both performance and speed. PFN-CEI has the best optimization performance, followed by PFN-Pen and GP-CEI+, with exceptional performance in engineering problems. With the unique transformer architecture and pre-trained nature, PFN-based BO shows its capability to accelerate the BO process by an order of magnitude compared to GP-based Bayesian optimization.

\section*{Declarations}
\subsection*{Funding} Cyril Picard: Swiss National Science Foundation (Postdoc.Mobility Fellowship P500PT\_206937)

\subsection*{Replication of results} This paper provides detailed descriptions of the benchmark example in the Appendix to support result replication. The Python code for the benchmarks and GP- and PFN-based BO  are provided in \url{https://github.com/rosenyu304/BOEngineeringBenchmark}. The GPs are built with the BoTorch library, and the PFN model is taken from ~\citet{muller2023pfns}. 

\subsection*{Conflict of Interest} On behalf of all authors, the corresponding author states that there is no conflict of interest.

\bibliography{sn-bibliography}%

\appendix
\begin{appendices}

\section{Benchmark Problems} \label{secA1}
\subsection{JLH1} This numerical problem is a two-dimensional ``sphere" problem featuring a single optimum and a continuous inequality constraint as detailed in~\citep{jetton2023constrained}. The domain of interest for both $x_1$ and $x_2$ is $[0,1]$.
\begin{gather*}
    f(x) = x_1^2 + x_2^2\\
    g(x) = x_1 + x_2 + 0.5 \leq 0
\end{gather*}

\subsection{JLH2} This numerical problem features a is two-dimensional objective with local optimum used in multiple research~\citep{campbell2024constraining, gardner2014bayesian}. The continuous inequality constraint is proposed by~\citep{campbell2024constraining}. The domain of interest is $x_1\in[-5,0]$ and $x_2\in[-5,5]$.  
\begin{gather*}
    f(x) = cos(2x_1)cos(x_2) + sin(x_1) \\
    g(x) = \frac{1}{4} (x_1+5)^2 
            + \frac{1}{100} x_2^2 - 1 \leq 0
\end{gather*}

\subsection{GKXWC1} The objective function of this question is identical to JHL2. However, the unique discontinuous inequality constraint is created by~\citep{gardner2014bayesian}. The domain of interest for both $x_1$ and $x_2$ is $[0,6]$.
\begin{gather*}
f(x) = cos(2x_1)cos(x_2) + sin(x_1) \\
g(x) =  cos(x_1)cos(x_2) - sin(x_1)sin(x_2) - 0.5 \leq 0
\end{gather*}

\subsection{GKXWC2}
This numerical problem is two-dimensional, featuring a multiple optimum and a discontinuous inequality constraint with a tiny feasible area proposed by~\citep{gardner2014bayesian}. The domain of interest for both $x_1$ and $x_2$ is $[0,6]$.
\begin{gather*}
    f(x) = sin(x_1) + y \\
    g(x) =  sin(x_1)sin(x_2) + 0.95 \leq 0
\end{gather*}

\subsection{Ackley Function} Ackley function is a popular scalable numerical test case for optimization. It has many local minimums with the global optimal at $x=[0,0,\cdots]^n$ for a $n-$dimensional problem. We consider the domain of interest to be $[-5,10]^n$ with $n=2,6,10$. The constraints formulation is taken from~\citep{eriksson2021scalable}. The large number of local minimum, multi-modal nature, and small feasible region makes this problem challenging.
\begin{gather*}
    f(x) = -20\exp{ \left(-0.2\sqrt{ \frac{1}{d} \sum^d_{i=1} x_i^2} \right) }  \\
        - \exp{ \left(\frac{1}{d} \sum^d_{i=1} cos(2\pi x_i) \right) } +20+\exp \left( 1 \right) 
        \\
    g_1(x) =  \sum^d_{i=1} x_i \leq 0 \\
    g_2(x) =  \lVert x \rVert_2 - 5 \leq 0
\end{gather*}

\subsection{Three Truss} The objective of this problem is to minimize the volume of the three-bar truss while each truss is constrained by the stress acting on it. The given information are length ($L$) is 100cm, pressure ($P$) is 2kN/$cm^2$, and stress ($\sigma$) is 2kN/$cm^2$. The domain of interest for both $x_1$ and $x_2$ is $[0,1]$. The formulation is taken from~\citep{yang2012bat}. 
\begin{gather*}
\qquad f(x) = ( 2\sqrt{2}x_1 + x_2 ) L
\\
\qquad g_1(x) = \frac{ (\sqrt{2}x_1 + x_2)P}{\sqrt{2}x_1^2 + 2x_1x_2} - \sigma
\\
\qquad g_2(x) = \frac{x_2 P}{\sqrt{2}x_1^2 + 2x_1x_2}  - \sigma
\\
\qquad g_3(x) = \frac{P}{ x_1 + \sqrt{2}x_2}  - \sigma
\end{gather*}

\subsection{Reinforced Concrete Beam} This 3D problem describes a beam supported at two end points spaced by 30 ft subjected to a live load and a dead load. The optimization objective is to minimize the cost while satisfying the constraints imposed by the loads and the safety requirement from ACI 318-77 code. This problem is representative of discrete value optimization. The cross-sectional area of the reinforcing bar ($A_s$) and the width of the concrete beam ($b$) are discrete while the depth of the concrete beam ($h$) is continuous. The domain of interest for each objective variable is: $A_s \in [0.2, 15]$ and $h\in[5,10]$. As for $b$, it is the integers in $[28,40]$. The formulation below is taken from~\citep{gandomi2011mixed}.
\begin{gather*}
\qquad f(x) = 29.4A_s + 0.6bh  
\\
\qquad g_1(x) = \frac{h}{b} - 4 \leq 0
\\
\qquad g_2(x) = 180 + \frac{7.35A_s^2}{b} - A_sh \leq 0
\end{gather*}

\subsection{Compression Spring} This optimal spring design problem has many variations in the literature. Here we pick the helical compression spring design optimization problem proposed by~\citep{gandomi2011mixed}. The goal is to minimize the spring volume. The optimization variables are: the number of spring coils (N), the winding coil diameter (D), and the wire diameter (d). The domain of interest for each objective variable is $d\in[0.05, 1]$, $D\in[0.25, 1.3]$, and $N\in[1.5, 2]$. 
\begin{gather*}
\qquad f(x) = (N+2)Dd^2 \\
\qquad g_1(x) = 1 -  \frac{D^3N} {71785 d^4 } \leq 0 \\
\qquad g_2(x) = \frac{4D^2-Dd} {12566(Dd^3-d^4)} + \frac{1}{5108d^2} -1\leq 0 \\
\qquad g_3(x) = 1 -  \frac{140.45d}{D^2N} \leq 0 \\
\qquad g_4(x) = \frac{D+d}{1.5} - 1\leq 0 \\
\end{gather*}

\subsection{Pressure Vessel} The optimization of a cylindrical pressure vessel with both ends capped is to minimize the cost while meeting the ASME constraints on boilers and pressure vessels. The objective (cost) is defined with four optimization variables (thickness of the cylindrical skin ($T_s$), thickness of the spherical head ($T_h$), the inner radius ($R$), and length of the cylindrical segment of the vessel ($L$)). The domain of interest for both $T_s$ and $T_h$ is $0.0625T$, where $T$ is a random integer value from $1\sim99$. For both $R$ and $L$, the domain of interest is $[10,200]$. The problem is originally proposed by~\citep{sandgren1990nonlinear} and the formulation is taken from~\citep{gandomi2011mixed}.
\begin{gather*}
\qquad f(x)= 0.6224T_s R L + 1.7781T_h R R 
\\+ 3.1661T_s T_s L + 19.84T_sT_s R  \\
\qquad g_1(x) = -T_s + 0.0193R \leq 0
\\
\qquad g_2(x) = -T_h + 0.00954R \leq 0
\\
\qquad g_3(x) = \pi R^2L - 4/3 \pi R^3 + 1296000 \leq 0
\\
\qquad g_4(x) = L-240 \leq 0
\end{gather*}

\subsection{Welded Beam} The welded beam is designed to minimize the manufacturing cost. The five constraints are imposed on the shear stress, bending stress in the beam, geometry, buckling load on the beam, and deflection of the beam. The optimization variables are the thickness of the weld ($h$), the length of the welded joint ($l$), the width of the beam ($t$), and the thickness of the beam ($b$). The domain of interest is $h\in[0.125,10], l\in[0.1,15], t\in[0.1,10]$, and $b\in[0.1,10]$. Details about the problem formulation can be found in~\citep{gandomi2011mixed}. 

\begin{gather*}
\qquad f(x) = 1.10471h^2l + 0.04811tb(14+l)
\\
\qquad \tau(x) = \sqrt{ \frac{ \tau'(x) ^2 + \tau''(x)^2 + l\tau'(x)\tau''(x)}{ \sqrt{0.25(l^2 + (h+t)^2)} } } 
\\
\qquad \tau'(x) = \frac{ 6000 }{ \sqrt{2}hl } 
\\
\qquad \tau''(x) = \frac{ 6000(14+0.5l) \sqrt{0.25(l^2 + (h+t)^2 )} }{ 2 (0.707hl  ( \frac{l^2}{12} + 0.25(h+t)^2 ) ) } 
\\
\qquad \sigma(x) = \frac{ 504000 }{ t^2 b } 
\\
\qquad P_c(x) = 64746(1-0.0282346t)tb^3
\\
\qquad \delta(x) = \frac{ 2.1952 }{ t^3 b } 
\\
\qquad g_1(x) = \tau(x) - 13600
\\
\qquad g_2(x) = \sigma(x) - 30000
\\
\qquad g_3(x) = b-h
\\
\qquad g_4(x) = P_c - 6000
\\
\qquad g_5(x) = 0.25 - \delta(x)
\end{gather*}

\subsection{Speed Reducer} The speed reducer design is one of the most famous benchmark problems in structural engineering. The objective is to minimize the weight of the speed reducer while considering the dimensions of the inner bearings and shafts. The domain of interest is $b\in[2.6,3.6], m\in[0.7,0.8], z\in[17,28], L_1$ and $L2\in[7.3, 8.3], d_1\in[2.9,3.9], $ and $ d_2\in[5,5.5]$. The problem originated from~\citep{golinski1973adaptive} and the formulation is taken from~\citep{yang2012bat}.
\begin{gather*}
\qquad f(x) = 0.7854bm^2 (3.3333z^2 + 14.9334z \\ - 43.0934)  -  1.508b(d_1^2 + d_2^2)
+ 7.4777(d_1^3 + d_2^3) \\ + 0.7854(L_1d_1^2 + L_2d_2^2)  )
\\
\qquad g_1(x) = \frac{27}{bm^2z} - 1 \leq 0
\\
\qquad g_2(x) = \frac{397.5}{bm^2z^2} -1 \leq 0
\\
\qquad g_3(x) = \frac{1.93L_1^3}{mzd_1^4} -1 \leq 0
\\
\qquad g_4(x) = \frac{1.93L_2^3}{mzd_2^4} -1 \leq 0
\\
\qquad g_5(x) = \frac{ \sqrt{ \left(\frac{745L_1}{mz} \right)^2 + 1.69\times10^6 } }{110d_1^3} -1 \leq 0
\\
\qquad g_6(x) = \frac{ \sqrt{ \left(\frac{745L_2}{mz} \right)^2 + 157.5\times10^6 } }{85d_2^3} -1 \leq 0
\\
\qquad g_7(x) = \frac{mz}{40} -1 \leq 0
\\
\qquad g_8(x) = \frac{5m}{B-1} -1 \leq 0
\\
\qquad g_9(x) = \frac{b}{12m} -1 \leq 0
\end{gather*}

\subsection{Heat Exchanger Design} The heat exchanger design is known for its six complicated binding constraints. Though the objective is simple, satisfying all linear ($g_1 \sim g_3$) and nonlinear constraints ($g_4 \sim g_6$) make it a difficult constrained optimization problem. The domain of interest is $x_1\in[10^2, 10^4]$, $x_{2\sim3}\in[10^3, 10^4]$, and  $x_{4\sim8}\in[10, 10^3]$. Details of the problem formulation can be found here~\citep{yang2012bat}.
\begin{gather*}
\qquad f(x) = x_1+x_2+x_3 \\
\qquad g_1(x) = 0.0025(x_4+x_6) - 1 \leq 0 \\
\qquad g_2(x) = 0.0025(x_5 + x_7 - x_4) - 1 \leq 0 \\
\qquad g_3(x) = 0.01(x_8-x_5) - 1 \\
\qquad g_4(x) = 833.33252x_4 + 100x_1 - x_1x_6\\ - 83333.333\leq 0 \\
\qquad g_5(x) = 1250x_5 + x_2x_4 - x_2x_7 - 125x_4 \leq 0 \\
\qquad g_6(x) = x_3x_5 - 2500x_5 - x_3x_8 + 1250000 \leq 0 \\
\end{gather*}

\subsection{Cantilever Stepped Beam} This problem is originally proposed to minimize the volume of the stepped cantilever beam by~\citep{thanedar1995survey}. In this 10D problem, we optimize a five-stepped cantilever beam with variable width $x_{i}$ and height $x_{i+5}$ of each step. The constant used are total length $L=100$, Young's modulus $E = 2 \times 107GPa$, and force applied on the end of the beam $P = 50000N$. The domain of interest for $x_{1\sim5}$ is $[1,5]$ and for $x_{6\sim10}$ is $[30,65]$. Details of the eleven constraints formulation can be found in ~\citep{yang2012bat, koziel2011computational}.
\begin{gather*}
    f(x) = \sum^5_{i=1} x_i x_{i+5} l_i \\
    g_1(x) = \frac{600P}{x_5x^2_{10}} - 14000 \leq 0 \\
    g_2(x) = \frac{6P(l_1+l_2)}{x_4 x^2_{9}} - 14000 \leq 0 \\
    g_3(x) = \frac{6P(l_1+l_2+l_3)}{x_3 x^2_{8}} - 14000 \leq 0 \\
    g_4(x) = \frac{6P(l_1+l_2+l_3+l_4)}{x_2 x^2_{7}} - 14000 \leq 0 \\
    g_5(x) = \frac{6P(l_1+l_2+l_3+l_4+l_5)}{x_1 x^2_{6}} - 14000 \leq 0 \\
    g_6(x) = \frac{Pl^3}{3E} \left( \frac{1}{l_5} + \frac{7}{l_4} + \frac{19}{l_3} + \frac{37}{l_2} + \frac{61}{l_1} \right) - 2.7 \leq 0 \\
    g_7(x) = \frac{x_{10}}{x_5} -20 \leq 0 \\
    g_8(x) = \frac{x_{9}}{x_4} -20 \leq 0 \\
    g_9(x) = \frac{x_{8}}{x_3} -20 \leq 0 \\
    g_{10}(x) = \frac{x_{7}}{x_2} -20 \leq 0 \\
    g_{11}(x) = \frac{x_{6}}{x_1} -20 \leq 0 \\
\end{gather*}

\subsection{Car Side Impact Design} The goal of this test case is to minimize the car weight while meeting the compatibility requirements for the car crash test. Eleven variables are used to describe the requirements imposed on the design, materials, and reinforcement of B-Pillar, floor side, cross members, door beam, door beltline, roof rail, barrier height, and impact position during the side impact test. This problem is originally proposed by~\citep{gu2001optimisation} and the formulation is taken from~\citep{gandomi2011mixed}. The domain for this test case is $x_{1, 3\sim7} \in [0.5, 1.5]$, $x_2\in[0.45, 1.35]$, $x_{8\sim9} \in [0.192, 0.345]$, and $x_{10\sim11}\in[-20, 0]$. 

\begin{gather*}
    \qquad \quad f(x) =1.98 + 4.90x_1 + 6.67x_2 + 6.98x_3 + \\
         4.01x_4 + 1.78x_5 + 2.73x_7 
    \\
    \qquad \quad g_1(x) = 1.16 - 0.3717x_2x_4 - 0.00931x_2x_{10} - \\ 0.484x_3x_9 + 0.01343x_6x_{10}   -1 \leq 0 
    \\
    g_2(x) = 0.261 - 0.0159x_1x_2 - 0.188x_1x_8 
    \\  - 0.019x_2x_7 + 0.0144x_3x_5 + 0.0008757x_5x_{10}
    \\  + 0.08045x_6x_9 + 0.00139x_8x_{11}  + 0.00001575x_{10}x_{11}    
    \\  -0.9 \leq 0 
    \\
    g_3(x) = 0.214 + 0.00817x_5 - 0.131x_1x_8 - 0.0704x_1x_9 
    \\  + 0.03099x_2x_6
              -0.018x_2x_7 + 0.0208x_3x_8 + 0.121x_3x_9 
    \\  - 0.00364x_5x_6
              +0.0007715x_5x_{10} - 0.0005354x_6x_{10} \\  +0.00121x_8x_{11}       -0.9 \leq 0
    \\
    g_4(x) = 0.74 -0.061x_2 -0.163x_3x_8 +0.001232x_3x_{10} \\ -0.166x_7x_9 +0.227x_2x_2        -0.9 \leq 0
    \\
    g_5(x) = 28.98 +3.818x_3-4.2x_1x_2+0.0207x_5x_{10}+ \\ 6.63x_6x_9-7.7x_7x_8+0.32x_9x_{10}  -32\leq 0
    \\
    g_6(x) = 33.86 +2.95x_3+0.1792x_{10}-5.057x_1x_2 \\ -11.0x_2x_8-0.0215x_5x_{10}-9.98x_7x_8+22.0x_8x_9 \\    -32 \leq 0
    \\
    g_7(x) = 46.36 -9.9x_2-12.9x_1x_8+0.1107x_3x_{10}  \\  -32\leq 0
    \\
    g_8(x) = 4.72 -0.5x_4-0.19x_2x_3 \\ -0.0122x_4x_{10} +0.009325x_6x_{10}+0.000191x_{11}^2 \\    -4\leq 0
    \\
    g_9(x) = 10.58 -0.674x_1x_2-1.95x_2x_8 \\ +0.02054x_3x_{10} -0.0198x_4x_{10}+0.028x_6x_{10} \\ -9.9\leq 0
    \\
    g_{10}(x) = 16.45 -0.489x_3x_7-0.843x_5x_6 \\ +0.0432x_9x_{10} -0.0556x_9x_{11}-0.000786x_{11}^2 \\ -15.7 \leq 0
\end{gather*}

\subsection{Keane Bump} The Keane Bump benchmark is a scalable optimization test case proposed by Keane~\citep{keane1994experiences} and it's known to be challenging for GP-based BO methods. The goal is to perform minimization over the domain $[0,10]^d$. Here we use $d=18$ since this is the maximum dimension that can be passed into the model used for PFNs4BO~\citep{muller2023pfns}. 
\begin{gather*}
    f(x) = -\left| \frac{ \sum^{d}_{i=1} cos^4(x_i) -2\prod_{i=1}^{d} cos^2(x_i)}{\sqrt{\sum^d_{i=1} ix_i^2}} \right| \\
    g_1(x) = 0.75-\prod_{i=1}^{d} x_i \leq 0 \\
    g_2(x) = \sum_{i=1}^{d} x_i - 7.5d \leq 0 \\
\end{gather*}

\end{appendices}

\end{document}